\PassOptionsToPackage{colorlinks}{hyperref}
\documentclass[aps,prd,preprint,superscriptaddress,nofootinbib,nobibnotes,longbibliography,12pt]{revtex4-2}
\usepackage[T1]{fontenc}
\usepackage{mathtools,amsmath,amsthm,amssymb}
\usepackage{calrsfs,dsfont,bbm,xcolor}
\usepackage{tikz,float}
\usetikzlibrary{shapes.geometric,arrows}
\tikzstyle{rect} = [draw,rectangle, fill=white!20, text width =3cm, text centered, minimum height = 1.5cm,scale=0.9]
\usepackage{url}
\DeclareMathAlphabet{\pazocal}{OMS}{zplm}{m}{n}
\usepackage{hyperref}
\hypersetup{
  colorlinks=true,
  citecolor=violet,
  linkcolor=black,
  urlcolor=black}
\begin{document}

\title{A more general exact solution for the Schwarzschild black hole in a Bertotti--Robinson--Bonnor--Melvin universe}
\author{Andrea Di Pinto}
\email{andrea.dipinto@mi.infn.it}
\affiliation{Dipartimento di Scienza e Alta Tecnologia (DISAT), Universit\`a degli Studi dell'Insubria, \\ Via Valleggio 11, I-22100 Como, Italy}
\affiliation{Istituto Nazionale di Fisica Nucleare (INFN), Sezione di Milano, \\ Via Celoria 16, I-20133 Milano, Italy}

\begin{abstract}
We begin by introducing a new parameterization for the Ovcharenko--Podolsk\'{y} family of exact solutions of the Einstein--Maxwell equations describing uncharged and non-accelerating static black holes immersed in an external Bertotti--Robinson electromagnetic field. The new spacetime is, in general, different from the original Ovcharenko--Podolsk\'{y} one, in particular on the branches solving the field equations, for which the original parameterization can lead to black holes with a negative mass, whereas the new one always leads to black holes with a positive mass. Moreover, by introducing additional gauge constants, we remove the various pathologies and ensure consistency with the laws of black hole thermodynamics. We then obtain a new family of Einstein--Maxwell solutions by generalizing this new parameterization, through Harrison transformation within the Ernst method, to also include an external electromagnetic field of the Bonnor--Melvin type. We again identify the appropriate gauge constants required to remove the possible pathologies and ensure that the resulting solutions satisfy the laws of black hole thermodynamics. Finally, we investigate the vacuum subcase of the new family and establish its relation to previously known metrics.
\end{abstract}
\maketitle

\newpage
\tableofcontents
\newpage
\hypersetup{urlcolor=blue,linkcolor=blue}
\section{Introduction}
\label{sec:introduction}
\noindent An increasing number of observations indicates that supermassive black holes at the centers of galaxies are surrounded by large-scale magnetic fields, sustained by currents in their surrounding magnetized accretion disks~\cite{Eatough:2013nva,EventHorizonTelescope:2021bee,You:2023dax,EventHorizonTelescope:2024hpu}. These fields are believed to play an important role in processes such as accretion, jet launching, and energy extraction. This provides a strong motivation for constructing exact solutions of the Einstein--Maxwell equations in General Relativity which describe black holes interacting with external electromagnetic fields. 

The most notable examples of analytical solutions representing universes permeated by an external electromagnetic field are certainly given by the Bonnor--Melvin spacetime~\cite{Bonnor:1954tis,Melvin:1963qx} and the Bertotti--Robinson one~\cite{Levi-Civita:2011ehb,Bertotti:1959pf,Robinson:1959ev}.
The construction of exact solutions describing black holes embedded in these external electromagnetic fields began with the works of Ernst and Wild~\cite{Ernst:1976mzr,Ernst:1976bsr}, who obtained black hole solutions in a Bonnor--Melvin spacetime (see also~\cite{DiPinto:2025yaa} for recent developments).

In a similar way, as regards black holes with a Bertotti--Robinson electromagnetic field, there are two main families of solutions: the one started by Alekseev and Garcia~\cite{Alekseev:1996fq,Alekseev:2025czq}, and the one found by Ovcharenko and Podolsk\'{y}~\cite{Podolsky:2025tle,Ovcharenko:2025cpm,Ovcharenko:2025qov,Ovcharenko:2026byw,Ovcharenko:2026pow}\footnote{To be more precise, the static subfamily of these metrics is equivalent to the solutions found by Van den Bergh and Carminati~\cite{VandenBergh:2020lvf}, which are, however, quite involved and purely mathematical, and whose physical meaning had never been clarified before the work of Ovcharenko and Podolsk\'{y}.}. However, at present, it remains unclear whether at least some static subcases of the two families are equivalent to each other.
Moreover, there is also the issue that, in the uncharged, non-accelerating, and static subcase, the Ovcharenko--Podolsk\'{y} family allows for \emph{two different} algebro-geometric structures of these black hole solutions, characterized by whether a certain parameter called $r_0$ is zero or not, while the Alekseev and Garcia solution has only one.

As one would expect, obtaining solutions describing black holes with both the Bonnor--Melvin and Bertotti--Robinson parameters has also started to gain interest, with most results being obtained as generalizations of the uncharged Ovcharenko--Podolsk\'{y} family: generalizations of the Schwarzschild black hole are found in~\cite{Astorino:2025lih, Astorino:2026okd}; of the accelerating Schwarzschild black hole, also called the C--metric, in~\cite{Barrientos:2026shy}; and of the Kerr black hole in~\cite{Ma:2026otg}.

In particular, the black hole found by Astorino in~\cite{Astorino:2025lih} corresponds to the solution obtained by adding the Bonnor--Melvin parameter to the $r_0=0$ Schwarzschild black hole in Bertotti--Robinson of the Ovcharenko--Podolsk\'{y} family, also called the the Schwarzchild$-\mathrm{BR}$ solution.
The aim of this paper is therefore to obtain the remaining static and uncharged black hole solution with both the Bonnor--Melvin and Bertotti--Robinson parameters, corresponding to the generalization of the $r_0\neq0$ Schwarzschild black hole in Bertotti--Robinson of the Ovcharenko--Podolsk\'{y} family, also referred to as the $\mathrm{RN}-\mathrm{BR}_0$ solution.

The paper is structured as follows. In section~\ref{sec:charged-static-family}, we provide a brief review of the Ovcharenko--Podolsk\'{y} static family of charged black holes accelerating in a Bertotti--Robinson background~\cite{Ovcharenko:2026byw}, focusing on the uncharged and non-accelerating subfamily and the issues it exhibits in certain subcases. In section~\ref{sec:new-uncharged-non-accelerating-subcase}, we first propose an alternative parameterization of the aforementioned subfamily that resolves these issues, and then proceed to also introduce and fix some gauge constants in this new paremeterization in order to remove the various singularities and to adjust the thermodynamic behavior. Moreover, we also study the equatorial geodesics for this family and find the explicit expression for the innermost stable circular orbit. Finally, in section~\ref{sec:bonnor-melvin-addition}, by means of the Harrison transformation of the Ernst method, we add the Bonnor--Melvin parameter into the newly found uncharged and non-accelerating parameterization, we again introduce and fix some gauge constants for the same reasons as discussed above, and at the end we study the vacuum subcase, as Astorino did for the $r_0 = 0$ branch in~\cite{Astorino:2026okd}. There, we show that the resulting metric in the $r_0 \neq 0$ branch is actually diffeomorphic to that of the $r_0 = 0$ branch, thereby rendering the $r_0$ parameter inessential in the vacuum subcase.

\section{Review of the Static Charged Family of Black Hole Solutions in Bertotti--Robinson}
\label{sec:charged-static-family}
In this section we write down the general static black hole solution recently found by Ovcharenko and Podolsk\'{y} in~\cite{Ovcharenko:2026byw}, describing an accelerating Reissner--Nordstr\"{o}m black hole embedded in a Bertotti--Robinson electromagnetic field, and then we derive and discuss the two subcases that arise in the non-accelerating and uncharged limit. Aside from the original paper, a review of the properties of some subcases of this class can also be found in~\cite{DiPinto:2026rvp}.

To fix notation, we begin by stating explicitly the Einstein--Maxwell equations adopted throughout this paper:
\begin{equation}
R_{\mu\nu} - \frac{1}{2} R\, g_{\mu\nu} =
2 \Bigl[ F_{\mu\rho} F_\nu{}^{\rho} - \frac{1}{4} F^2 g_{\mu\nu} \Bigr] \,, \quad
\nabla_\mu F^{\mu\nu} = 0 \,, \label{einstein-maxwell}
\end{equation}
where $F_{\mu \nu} = \partial_{\mu}A_{\nu}-\partial_{\nu}A_{\mu}$ is the Faraday tensor of the electromagnetic field.

The metric and the $1-$form solving the above Einstein--Maxwell field equations are given by
\begin{subequations}
\label{accelerating-rn-br}
\begin{align}
{ds}^2 & = \frac{1}{\Omega^2}
\biggl[-\frac{\Delta_r} {r^2}{dt}^2 + r^2\biggl(\frac{{dr}^2}{\Delta_r}
+ \frac{{d\theta}^2}{\Delta_\theta}\biggr) + r^2\sin^2\theta\Delta_\theta {d\phi}^2\biggr] \,, \label{accelerating-rn-br-metric} \\
A & = \frac{w}{B} \frac{\partial_\theta\Omega}{r\sin\theta} dt
+ \frac{\sqrt{1-w^2}}{B} \bigl[\Omega - 1 - r\,\partial_r\Omega\bigr] d\phi \,, \label{accelerating-rn-br-potential}
\end{align}
\end{subequations}
where
\begin{subequations}
\label{accelerating-rn-br-functions}
\begin{align}
\Delta_r & = \bigl[(r-r_0)^2(1-B^2 m^2)- 2 m (r-r_0)\bigr]\bigl[1+(B^2-\alpha^2)(r-r_0)^2\bigr] \,, \\
\Delta_\theta & = 1 - 2\alpha m \cos\theta + B^2 m^2 \cos^2\theta \,, \\
\begin{split}
\Omega^2 & = \bigl[1-\alpha(r-r_0)\cos\theta\bigl]^2 +\,\\
 &\quad + B^2 \bigl[(r-r_0)^2(\sin^2\theta+B^2m^2\cos^2\theta-2\alpha m \cos\theta)+ 2 m (r-r_0)\cos^2\theta\bigl] \,.
\end{split}
\end{align}
\end{subequations}
The parameter $r_0$ cannot assume an arbitrary value: instead, only two distinct \emph{discrete} possibilities are allowed\footnote{These two discrete branches arise from imposing the absence of twist on the more general class of charged and rotating black holes immersed in a Bertotti--Robinson background~\cite{Ovcharenko:2025cpm}.}:
\begin{align}
r_0 & = 0 \,, \label{r0-zero-charged-a} \tag{case I-a}\\
r_0 & = \frac{2 m B^2}{\alpha^2-B^2(1-B^2 m^2)} \,. \label{r0-not-zero-charged-a} \tag{case II-a}
\end{align}
The physical meaning of the parameters is as follows: $m$ is related to the mass of the black hole, $B$ determines the value of the Bertotti--Robinson electromagnetic field, $\alpha$ corresponds to the acceleration, and $w$ is the electromagnetic duality parameter, in the sense that the electric charge $\mathrm{Q}$ and the magnetic charge $\mathrm{P}$ are given by\footnote{We recall that ${}^\star F^{\mu\nu} = \frac{1}{2\sqrt{-g}}\varepsilon^{\mu \nu \rho \sigma} F_{\rho \sigma}$ is the Hodge dual of the Faraday tensor $F^{\mu \nu}$.}
\begin{subequations}
\label{charges-general}
\begin{align}
\mathrm{Q} & = -\frac{1}{8\pi}\oint_\Sigma F^{\mu\nu} d\Sigma_{\mu \nu} = \frac{\alpha\,r_0}{B} w \,, \\
\mathrm{P} & = -\frac{1}{8\pi}\oint_\Sigma {}^\star F^{\mu\nu} d\Sigma_{\mu \nu} = \frac{\alpha\,r_0}{B} \sqrt{1-w^2} \,,
\end{align}
\end{subequations}
meaning that the solution is fully electric for $w=1$, while it is fully magnetic for $w=0$.

As explained in~\cite{Ovcharenko:2026byw}, this class of solutions~\eqref{accelerating-rn-br} is actually equivalent to the one found by Van den Bergh and Carminati~\cite{VandenBergh:2020lvf}, who obtained all non-twisting and non-aligned Einstein--Maxwell solutions of Petrov type D within the Robinson--Trautman class.

In this way, the physical difference between the two branches becomes clear, in the sense that the $r_0 = 0$ branch~\eqref{r0-zero-charged-a} corresponds to the $hj=0$ subcase of the Van den Bergh--Carminati, for which the aligned component of the electromagnetic field is determined by the non-aligned one. Conversely, the $r_0 \neq 0$ branch~\eqref{r0-not-zero-charged-a} is equivalent to the $hj \neq 0$ subcase of Van den Bergh--Carminati, in which the aligned and non-aligned components are mutually independent.
\subsection{The Uncharged and Non-Accelerating Subcase}
\label{sec:uncharged-non-accelerating-subcase}
The requirement for the black hole to be uncharged, $\mathrm{Q}=\mathrm{P}=0$~\eqref{charges-general}, is automatically satisfied in the $r_0 = 0$ branch \eqref{r0-zero-charged-a}, while for the $r_0 \neq 0$ branch \eqref{r0-not-zero-charged-a}, the absence of acceleration, $\alpha = 0$, is required. In other words, if we remove the acceleration from the general static solution~\eqref{accelerating-rn-br}, we directly obtain the general subfamily of uncharged and non-accelerating black holes in a Bertotti--Robinson background, which is then given by:
\begin{subequations}
\label{NOT-accelerating-rn-br}
\begin{align}
{ds}^2 & = \frac{1}{\Omega^2}
\biggl[-\frac{\Delta_r} {r^2}{dt}^2 + r^2\biggl(\frac{{dr}^2}{\Delta_r}
+ \frac{{d\theta}^2}{\Delta_\theta}\biggr) + r^2\sin^2\theta\Delta_\theta {d\phi}^2\biggr] \,, \label{NOT-accelerating-rn-br-metric} \\
A & = \frac{w}{B} \frac{\partial_\theta\Omega}{r\sin\theta} dt
+ \frac{\sqrt{1-w^2}}{B} \bigl[\Omega - 1 - r\,\partial_r\Omega\bigr] d\phi \,, \label{NOT-accelerating-rn-br-potential}
\end{align}
\end{subequations}
with
\begin{subequations}
\label{NOT-accelerating-rn-br-functions}
\begin{align}
\Delta_\theta & = 1+ B^2 m^2 \cos^2\theta \,, \\
\Delta_r & = \bigl[(r-r_0)^2(1-B^2 m^2)- 2 m (r-r_0)\bigr]\bigl[1+B^2(r-r_0)^2\bigr] \,, \\
\Omega^2 & = 1+ B^2 \bigl[(r-r_0)^2(\sin^2\theta+B^2m^2 \cos^2\theta)+ 2 m (r-r_0)\cos^2\theta\bigl] \,,
\end{align}
\end{subequations}
and where now the two possible $r_0$ branches are given by:
\begin{align}
r_0 & = 0 \,, \label{r0-zero-not-charged-b} \tag{case I-b}\\
r_0 & = -\frac{2 m}{1-B^2 m^2} \,. \label{r0-not-zero-not-charged-b} \tag{case II-b}
\end{align}
The $r_0 = 0$ subcase \eqref{r0-zero-not-charged-b} is called the Schwarzschild black hole in Bertotti--Robinson, or also just the Schwarzschild$-\mathrm{BR}$ solution, since in the $B\to0$ limit it automatically reduces to the Schwarzschild solution, which represents a black hole of mass $m$:
\begin{equation}
\label{positive-Schwarzschild}
{ds}^2\Big\rvert_{\bigl(\alpha=B=0,\,\,\, r_0=0\bigr)} =-\biggl(1-\frac{2m}{r}\biggr){dt}^2 + \frac{{dr}^2}{\bigl(1-\frac{2m}{r}\bigr)}
+ r^2{d\theta}^2 + r^2\sin^2\theta\, {d\phi}^2 \,.
\end{equation}
In a similar way, the $r_0 \neq 0$ subcase \eqref{r0-not-zero-not-charged-b} is referred to as the $\mathrm{RN}-\mathrm{BR}_0$ solution, because it is obtained as an uncharged subcase of the Reissner--Nordstr\"{o}m black hole in a Bertotti--Robinson background, yet distinct from the Schwarzschild$-\mathrm{BR}$ solution. However, the main problem with this solution is that it still reduces to Schwarzschild in the $B\to0$ limit, yet with a \emph{negative} mass $\tilde{m}=-m<0$:
\begin{align}
\label{negative-Schwarzschild}
\begin{split}
{ds}^2\Big\rvert_{\bigl(\alpha=B=0,\,\,\, r_0\neq0\bigr)} & = -\biggl(1+\frac{2m}{r}\biggr){dt}^2 + \frac{{dr}^2}{\bigl(1+\frac{2m}{r}\bigr)} + r^2{d\theta}^2 + r^2\sin^2\theta\, {d\phi}^2 \,, \\
& = -\biggl(1-\frac{2\tilde{m}}{r}\biggr){dt}^2 + \frac{{dr}^2}{\bigl(1-\frac{2\tilde{m}}{r}\bigr)} + r^2{d\theta}^2 + r^2\sin^2\theta\, {d\phi}^2 \,.
\end{split}
\end{align}
\vspace{0.25cm}
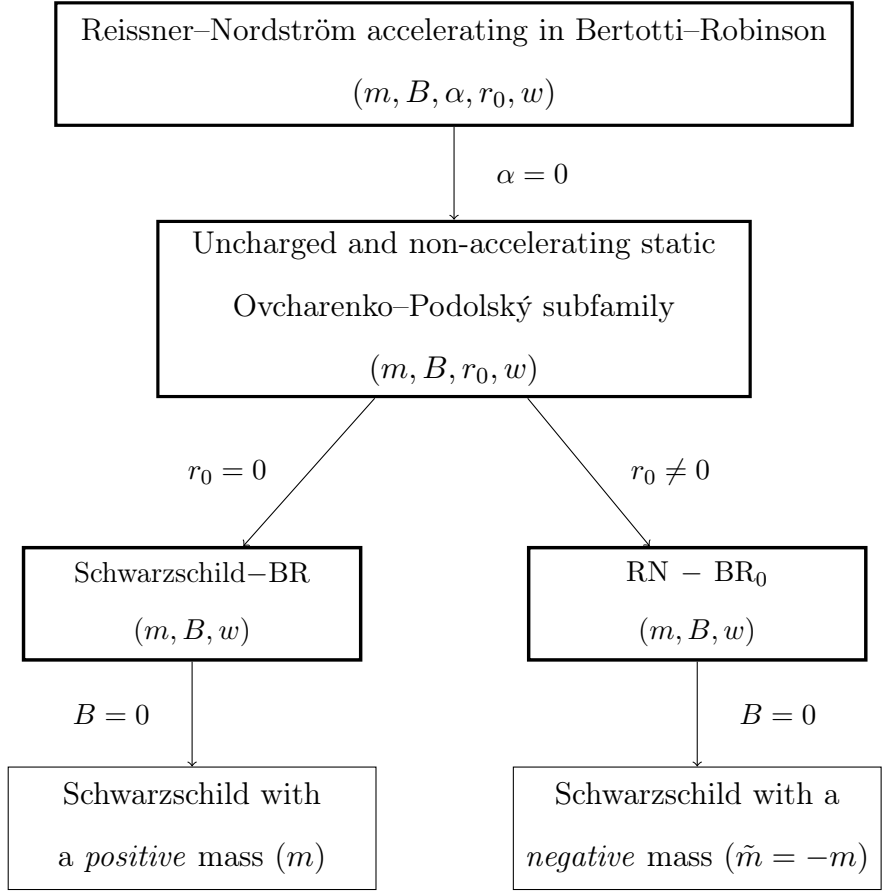
\begin{figure}[H]
\begin{center}
\begin{tikzpicture}
\node[rect,scale=1.2,text width=9.5cm,line width=1.2pt](RN-BR){Reissner--Nordstr\"{o}m accelerating in Bertotti--Robinson \\ ($m,B,\alpha,r_0,w$)};
    \node[rect, scale=1.2,anchor=north,
    text width=7cm,below of=RN-BR,node distance=3cm,line width=1.2pt](Uncharged){Uncharged and non-accelerating static Ovcharenko--Podolsk\'{y} subfamily \\ ($m,B,r_0,w$)};
        \node[scale=1.2,anchor=north,
    text width=4.5cm,below of=Uncharged,node distance=3.25cm](empty){};
        \node[rect,scale=1.1,anchor=north,text width=4.25cm, left of=empty, node distance=3.5cm,line width=1.3pt](S-BR){Schwarzschild$-\mathrm{BR}$ \\ ($m,B,w$)};
        \node[rect,scale=1.1,anchor=north,text width=4.25cm, right of=empty, node distance=3.25cm,line width=1.3pt](RN-BR0){$\mathrm{RN}-\mathrm{BR}_0$\\ ($m, B, w$)};
            \node[rect,scale=1.2,anchor=north, below of=S-BR, text width=4.25cm, node distance=2.75cm](S){Schwarzschild with a \emph{positive} mass ($m$)};
            \node[rect,scale=1.2,anchor=north,text width=4.25cm, below of=RN-BR0, node distance=2.75cm](S-Minus){Schwarzschild with a \emph{negative} mass ($\tilde{m}=-m$)};

\draw[->] (RN-BR) -- node [right] {\; \, $\alpha = 0$}(Uncharged);
\draw[->] (Uncharged) -- node [left] {$r_0 = 0$  \; \,}(S-BR);
\draw[->] (Uncharged) -- node [right] { \; \, $r_0 \neq 0$}(RN-BR0);
\draw[->] (S-BR) -- node [left] {$B=0$ \; \,} (S);
\draw[->] (RN-BR0) -- node [right] { \; \, $B=0$} (S-Minus);
\end{tikzpicture}
\caption{Diagram of the limiting subcases of the general accelerating Reissner--Nordstr\"om black hole embedded in a Bertotti--Robinson background~\cite{Ovcharenko:2026byw}, obtained by successively setting $\alpha=0$, $r_0=0$ (or $r_0\neq0$), and $B=0$. As depicted, both the $r_0=0$ and $r_0\neq0$ branches reduce to the Schwarzschild solution in the $B\to0$ limit; however, in the $r_0\neq0$ case the resulting black hole has a \emph{negative} mass.}
\label{fig:Graph-1}
\end{center}
\end{figure}
\section{A Single Physical Metric for the Uncharged Black Holes in Bertotti--Robinson}
\label{sec:new-uncharged-non-accelerating-subcase}
As depicted in figure~\ref{fig:Graph-1}, starting from the charged Reissner--Nordstr\"{o}m black hole accelerating in Bertotti--Robinson~\eqref{accelerating-rn-br}, one can obtain an uncharged and non-accelerating static subfamily of black holes embedded in the same background~\eqref{NOT-accelerating-rn-br}; this subfamily possesses two branches, corresponding to the two possible values of the parameter $r_0$. The $r_0 = 0$ branch~\eqref{r0-zero-not-charged-b} reduces, in the $B\to 0$ limit, to a physical Schwarzschild black hole with \emph{positive} mass, whereas the $r_0 \neq 0$ branch~\eqref{r0-not-zero-not-charged-b} reduces, in the same limit, to a Schwarzschild solution with \emph{negative} mass.

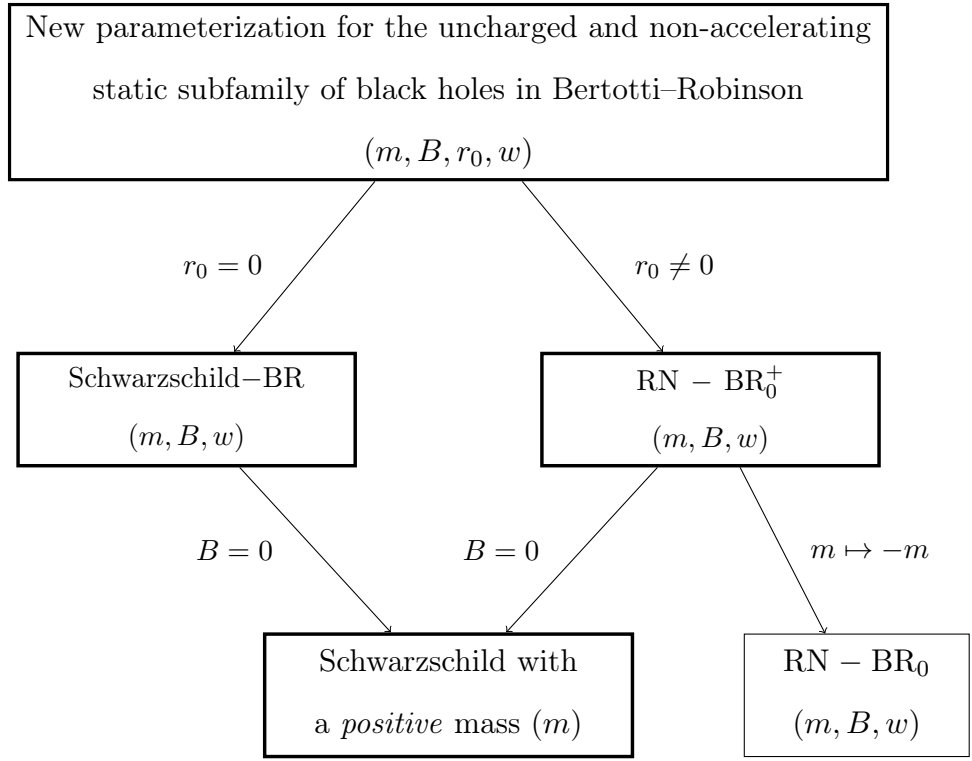
\begin{figure}[H]
\vspace{0.5cm}
\begin{center}
\begin{tikzpicture}
\node[rect, scale=1.2,
    text width=10.5cm,line width=1.2pt](Uncharged-New){New parameterization for the uncharged and non-accelerating static subfamily of black holes in Bertotti--Robinson\\ ($m,B,r_0, w$)};
    \node[scale=1.2,anchor=north,
    text width=4.5cm,below of=Uncharged-New,node distance=3.5cm](empty){};
    \node[rect,scale=1.1,anchor=north,text width=4.25cm, left of=empty, node distance=3.5cm,line width=1.3pt](S-BR){Schwarzschild$-\mathrm{BR}$ \\ ($m,B, w$)};
    \node[rect,scale=1.1,anchor=north,text width=4.25cm, right of=empty, node distance=3.5cm,line width=1.3pt](RN-BR0-Plus){$\mathrm{RN}-\mathrm{BR}_0^{+}$\\ ($m, B, w$)};
        \node[rect,scale=1.2,anchor=north, below of=empty, text width=4.25cm, node distance=3.5cm,line width=1.3pt](S){Schwarzschild with a \emph{positive} mass ($m$)};
        \node[rect,scale=1.2,anchor=north, right of=S, text width=2.5cm, node distance=5cm](RN-BR0){$\mathrm{RN}-\mathrm{BR}_0$\\ ($m, B, w$)};

\draw[->] (Uncharged-New) -- node [left] {$r_0 = 0$ \; \,}(S-BR);
\draw[->] (Uncharged-New) -- node [right] {\; \, $r_0 \neq 0$}(RN-BR0-Plus);
\draw[->] (S-BR) -- node [left] {$B=0$ \; \,} (S);
\draw[->] (RN-BR0-Plus) -- node [left] {$B=0$ \; \, } (S);
\draw[->] (RN-BR0-Plus) -- node [right] {\; $m\mapsto - m$} (RN-BR0);
\end{tikzpicture}
\vspace{0.1cm}
\caption{Diagram of the limiting subcases of the new, different parameterization given in this section for the uncharged and non-accelerating static subfamily of black holes embedded in a Bertotti--Robinson background. As depicted, both the $r_0=0$ and $r_0\neq0$ branches now reduce to the Schwarzschild solution with a \emph{positive} mass in the $B\to0$ limit.}
\label{fig:Graph-2}
\end{center}
\end{figure}

Therefore, we provide here a new and different parameterization for the uncharged and non-accelerating subclass of black holes, characterized by the fact that now both branches reduce to the Schwarzschild solution with \emph{positive} mass in the $B \to 0$ limit:
\begin{subequations}
\label{New-NOT-accelerating-rn-br}
\begin{align}
{ds}^2 & = \frac{1}{\Omega^2}
\biggl[-\frac{\Delta_r} {r^2}{dt}^2 + r^2\biggl(\frac{{dr}^2}{\Delta_r}
+ \frac{{d\theta}^2}{\Delta_\theta}\biggr) + r^2\sin^2\theta\Delta_\theta {d\phi}^2\biggr] \,, \label{New-NOT-accelerating-rn-br-metric} \\
A & = \frac{w}{B} \frac{\partial_\theta\Omega}{r\sin\theta} dt
+ \frac{\sqrt{1-w^2}}{B} \bigl[\Omega - 1 - r\,\partial_r\Omega\bigr] d\phi \,, \label{New-NOT-accelerating-rn-br-potential}
\end{align}
\end{subequations}
with
\begin{subequations}
\label{New-NOT-accelerating-rn-br-functions}
\begin{align}
\Delta_\theta & = 1 + B^2 m^2 \cos^2\theta \,, \\
\Delta_r & = \bigl[r^2(1-B^2 m^2)- 2 m r\bigr]\bigl[1+B^2(r-r_0)^2\bigr] \,, \\
\Omega^2 & = \bigl[1+ B^2(r-r_0)^2\bigr]- \bigl[r^2(1-B^2 m^2)- 2 m r\bigr]B^2\cos^2\theta \,,
\end{align}
\end{subequations}
where the two branches are now given by:
\begin{align}
r_0 & = 0 \,, \label{r0-zero-not-charged-c} \tag{case I-c}\\
r_0 & = \frac{2 m}{1-B^2 m^2} \,. \label{r0-not-zero-not-charged-c} \tag{case II-c}
\end{align}
This new parameterization has been obtained by suitably modifying the Ovcharenko--Podolsk\'{y} uncharged subfamily~\eqref{NOT-accelerating-rn-br}, namely by adding additional terms proportional to $r_0$ in such a way that the $r_0=0$ branch remains unchanged while the sign of the $m$ parameter is flipped in the $r_0  = \frac{2 m}{1-B^2 m^2}$ branch.
Indeed, as can be easily verified and as depicted in figure~\ref{fig:Graph-2}, the $r_0 = 0$ branch~\eqref{r0-zero-not-charged-c} still corresponds to the Schwarzschild$-\mathrm{BR}$ solution. On the other hand, for general $r_0 \neq 0$, it differs from the Ovcharenko--Podolsk\'{y} solution~\eqref{NOT-accelerating-rn-br}. Nevertheless, in the $r_0 \neq 0$ branch~\eqref{r0-not-zero-not-charged-c}, which solves the field equations~\eqref{einstein-maxwell}, it is related to the $\mathrm{RN}-\mathrm{BR}_0$ solution by a change of sign of the mass parameter, $m \mapsto -m$. For this reason, we refer to this spacetime as the $\mathrm{RN}-\mathrm{BR}_0^{+}$ solution. Obviously, given the explicit relation to the previous subfamily, it is also straightforward that both branches reduce to Schwarzschild with a \emph{positive} mass in the $B \to 0$ limit. 

Moreover, in this new parameterization, both $r_0$ branches possess a single black hole horizon, corresponding to the outer zero of the $\Delta_{r}$ function~\eqref{New-NOT-accelerating-rn-br-functions}, located at the same position for both solutions, independent of the parameter $r_0$:
\begin{equation}
\label{horizon-New-NOT-accelerating-rn-br}
r_H=\frac{2 m}{1-B^2 m^2} \,,
\end{equation}
which, in the physical range of the parameters, i.e.~$m>0$ and $|B|<\frac{1}{m}$, is always greater than the corresponding Schwarzschild horizon $r_{H-\mathrm{Schwarzschild}} = 2m < r_H$.
\subsection{Regularized Form of the New Metric}
\label{sec:fixed-new-uncharged-non-accelerating-subcase}
Since the metric found above may in general exhibit pathologies, we now introduce additional constants into the metric and electromagnetic potential, namely $\delta_{t}$, $C_f$, $\delta_{\phi}$, and $\delta A_{\phi}$, so that they can be suitably employed to remove such pathologies if needed, as follows:
\begin{subequations}
\label{New-semi-fixed-NOT-accelerating-rn-br}
\begin{align}
{ds}^2 & = \frac{1}{\Omega^2}
\biggl[-\frac{\Delta_r}{r^2}\delta^2_{t}\,{dt}^2 + C^2_f\,r^2\biggl(\frac{{dr}^2}{\Delta_r}
+ \frac{{d\theta}^2}{\Delta_\theta}\biggr) + r^2\sin^2\theta\,\Delta_\theta\, \delta^2_{\phi}\, {d\phi}^2\biggr] \,, \label{New-semi-fixed-NOT-accelerating-rn-br-metric} \\
A & = \frac{w}{B} \frac{\partial_\theta\Omega}{r\sin\theta}\,\delta_t\, dt
+ \frac{\sqrt{1-w^2}}{B} \bigl[\Omega - 1 - r\,\partial_r\Omega \bigr]\delta_{\phi}\, d\phi - \delta A_\phi\,\delta_{\phi}\, d\phi\,, \label{New-semi-fixed-NOT-accelerating-rn-br-potential}
\end{align}
\end{subequations}
while the remaining functions are unchanged and retain the form given in equations~\eqref{New-NOT-accelerating-rn-br-functions}.

Clearly, the constants $\delta_{t}$ and $\delta_{\phi}$ can be reabsorbed by means of a simple coordinate transformation, $\delta A_{\phi}$ similarly by a gauge transformation, and $C_f$ by an overall constant rescaling of the metric. Nevertheless, these constants give rise not only to topological effects, but also to different thermodynamic properties, as we will see below. We therefore now proceed to compute the quantities associated with the various possible pathologies, and to verify whether this metric satisfies the laws of black hole thermodynamics.
\subsubsection{Conical Singularities}
\label{sec:fixed-new-uncharged-non-accelerating-subcase-conical-singularities}
Axial conical singularities constitute a type of spacetime pathology arising from a defect in the azimuthal angle, resulting in an irregular symmetry axis. The angular defects can be computed as the ratio between the length of a small circle surrounding the azimuthal axis and its radius~\cite{DiPinto:2025yaa}, evaluated on both halves of the symmetry axis, $\theta=0$ and $\theta=\pi$, since the angular deficit can in general differ between the two halves:
\begin{equation}
\label{conical-condition-0}
\delta_0 = \lim_{\theta \to 0} \frac{1}{\theta} \int_{0}^{2 \pi} \!\!\!\sqrt{\frac{g_{\phi\phi}}{g_{\theta\theta}}}\,d\phi \,, \quad \quad \quad
\delta_\pi =  \lim_{\theta \to \pi} \frac{1}{\pi-\theta} \int_{0}^{2 \pi} \!\!\!\sqrt{\frac{g_{\phi\phi}}{g_{\theta\theta}}}\,d\phi \,.
\end{equation}
A spacetime is thus free of conical singularities provided that
\begin{equation}
\label{conical-condition}
\delta_0 = \delta_\pi = 2\pi \,.
\end{equation}
Applying this to the spacetime under consideration~\eqref{New-semi-fixed-NOT-accelerating-rn-br}, we find:
\begin{equation}
\label{New-semi-fixed-NOT-accelerating-rn-br-conical-condition-0}
\delta_0 = \delta_\pi =  \frac{2 \pi(1+B^2m^2) \delta_\phi}{C_f} \,,
\end{equation}
meaning that the absence of conical singularities is achieved provided that
\begin{equation}
\label{New-semi-fixed-NOT-accelerating-rn-br-conical-condition}
\delta_{\phi} = \frac{C_f}{(1+B^2m^2)}  \,.
\end{equation}
\subsubsection{Dirac Strings}
\label{sec:fixed-new-uncharged-non-accelerating-subcase-dirac-strings}
A magnetic charge that cannot be described by a smooth magnetic potential results in a discontinuity that generates a line singularity, called the Dirac string~\cite{Dirac:1931kp,Dirac:1948um}. These strings can be interpreted as one-dimensional curves in space that connect two Dirac monopoles with opposite magnetic charges, or, similarly, extend from a single magnetic monopole out to infinity. The condition ensuring that a Maxwell potential $A_{\mu}$ does \emph{not} exhibit Dirac strings is:
\begin{equation}
\label{dirac-condition}
\lim_{\theta \to 0} A_{\phi} = \lim_{\theta \to \pi} A_{\phi} = 0 \,.
\end{equation}
For our case~\eqref{New-semi-fixed-NOT-accelerating-rn-br}, this condition turns out to be satisfied whenever
\begin{equation}
\label{New-semi-fixed-NOT-accelerating-rn-br-diract-condition}
\delta A_\phi =  \sqrt{1-w^2}\,B\, m\, r_0\,.
\end{equation}
In particular, we observe that the addition of this gauge constant is required only in the $r_0\neq0$ branch~\eqref{r0-not-zero-not-charged-c}.
\subsubsection{Misner Strings and Closed Timelike Curves}
\label{sec:fixed-new-uncharged-non-accelerating-subcase-misner-strings-and-ctcs}
Misner strings~\cite{Misner:1963fr,Astefanesei:2004kn}, the gravitational counterpart of Dirac strings, arise whenever the off-diagonal function $g_{t\phi}/g_{tt}$ fails to be regular on the symmetry axis~\cite{Alekseev:2019kcf}, assuming different values on the two hemispheres $\theta=0$ and $\theta=\pi$:
\begin{equation}
\lim_{\theta \to 0} \frac{g_{t\phi}}{g_{tt}} \neq \lim_{\theta \to \pi} \frac{g_{t\phi}}{g_{tt}} \,. \label{misner-condition}
\end{equation}
Closed timelike curves (CTCs) are timelike curves that close back on themselves, thereby violating causality. For an axisymmetric spacetime endowed with a Killing vector $\zeta=\partial_{\phi}$ whose orbits are closed, the presence of CTCs is determined by the region where
\begin{equation}
\zeta^2 = g_{\mu \nu} \zeta^{\mu}\zeta^{\nu} = g_{\phi \phi} < 0 \,. \label{ctcs-condition}
\end{equation}
Given that the solution under consideration~\eqref{New-semi-fixed-NOT-accelerating-rn-br} is non-rotating, namely it has a diagonal metric, and that the component $g_{\phi\phi}$ is always positive, these two pathologies can never occur.
\subsubsection{Mass and Thermodynamics}
\label{sec:fixed-new-uncharged-non-accelerating-subcase-thermodynamics}
Since the spacetime under study is not asymptotically flat, it is not guaranteed that the black hole mass corresponds to the one obtained using the usual Komar prescription~\cite{Komar:1958wp}. Indeed, proceeding in this way yields a complex result that is not physically well defined. Nevertheless, we continue to compute the black hole mass $\mathrm{M}$ as a Komar integral, but instead of taking the integration surface at infinity, we evaluate the integral on the black hole horizon, $r_H=\frac{2m}{1-B^2m^2}$~\eqref{horizon-New-NOT-accelerating-rn-br}, and subsequently verify whether the resulting expression is consistent with black hole thermodynamics.
\begin{equation}
\mathrm{M}=-\frac{1}{4\pi}\oint_{\Sigma_{H}} \nabla^{\mu}\xi^{\nu}\, d\Sigma_{\mu\nu} = m \,\delta_{t}\,\delta_{\phi} \label{komar-mass} \,,
\end{equation}
where $\xi=\partial_t$ is the timelike Killing vector, $d\Sigma_{\mu\nu} = -n_{[\mu}\sigma_{\nu]}\sqrt{g_{\theta\theta}g_{\phi\phi}}\,d\theta d\phi$ denotes the infinitesimal integration surface element, with $n_{\mu}$ and $\sigma_{\nu}$ being the orthonormal timelike and spacelike vectors normal to the integration surface $\Sigma_{H}$ evaluated at the horizon.

The surface gravity $\kappa$~\cite{Bardeen:1973gs} and the temperature $\mathrm{T}$ at the horizon $r_H$ are given by:
\begin{subequations}
\label{surface-gravity-temperature-New-NOT-accelerating-rn-br}
\begin{align}
\kappa = \sqrt{-\frac{1}{2} \nabla_{\mu}\xi_{\nu}\nabla^{\mu}\xi^{\nu}}\Bigg\rvert_{r=r_{H}} \!& = \frac{\bigl[(1+B^2 m^2)^2-2 B^2 m (1-B^2 m^2)r_0\bigr]\delta_{t}}{4\,C_f\,m}\,, \label{surface-gravity-New-NOT-accelerating-rn-br}\\
\mathrm{T} = \frac{\kappa}{2\pi} & = \frac{\bigl[(1+B^2 m^2)^2-2 B^2 m (1-B^2 m^2)r_0\bigr]\delta_{t}}{8\pi\,C_f\,m} \,. \label{temperature-New-NOT-accelerating-rn-br}
\end{align}
\end{subequations}
The horizon area $\mathrm{A}$ and the Bekenstein--Hawking entropy $\mathrm{S}$~\cite{Bekenstein:1973ur,Hawking:1975vcx} are
\begin{subequations}
\label{area-entropy-New-NOT-accelerating-rn-br}
\begin{align}
\mathrm{A} = \int_{0}^{2 \pi}d\phi \int_{0}^{\pi} \sqrt{g_{\phi \phi}g_{\theta\theta}}\,d\theta\Bigg\rvert_{r=r_{H}} \! & = \frac{16 \pi\, m^2\,C_f\,\delta_{\phi}}{\bigl[(1+B^2 m^2)^2-2 B^2 m (1-B^2 m^2)r_0\bigr]} \,, \label{area-New-NOT-accelerating-rn-br}\\
\mathrm{S} = \frac{\mathrm{A}}{4} & = \frac{4 \pi\, m^2\,C_f\,\delta_{\phi}}{\bigl[(1+B^2 m^2)^2-2 B^2 m (1-B^2 m^2)r_0\bigr]}\,. \label{entropy-New-NOT-accelerating-rn-br}
\end{align}
\end{subequations}
From this, we immediately find that the Smarr law~\cite{Smarr:1972kt} holds
\begin{subequations}
\label{smarr-law-New-NOT-accelerating-rn-br}
\begin{align}
\mathrm{M}= 2\,\mathrm{T}\,\mathrm{S} = m \,\delta_{t}\,\delta_{\phi} \,,
\end{align}
\end{subequations}
for any values of the gauge constants $\delta_{t}$ and $\delta_{\phi}$.

On the other hand, the Christodoulou--Ruffini mass formula~\cite{Christodoulou:1971pcn}
\begin{equation}
\mathrm{M}^2 = \frac{\mathrm{S}}{4\pi} \,, \label{Ruffini-general}
\end{equation}
turns out to be satisfied only if
\begin{equation}
C_f = \bigl[(1+B^2 m^2)^2-2 B^2 m (1-B^2 m^2)r_0\bigr]\delta^2_{t} \delta_{\phi} \label{Ruffini-New-NOT-accelerating-rn-br-condition}\,.
\end{equation}
This result has an important implication, because by substituting it into the previous expressions for the temperature~\eqref{temperature-New-NOT-accelerating-rn-br} and the entropy~\eqref{entropy-New-NOT-accelerating-rn-br}, we obtain:
\begin{subequations}
\label{temperature-entropy-Fixed-New-NOT-accelerating-rn-br}
\begin{align}
\mathrm{T} & = \frac{1}{8\pi\,\mathrm{M}} \,, \label{temperature-Fixed-New-NOT-accelerating-rn-br}\\
\mathrm{S} & = 4\pi\,\mathrm{M}^2 \,, \label{entropy-Fixed-New-NOT-accelerating-rn-br}
\end{align}
\end{subequations}
which is exactly what one obtains for the Schwarzschild solution, in the absence of the Bertotti--Robinson electromagnetic field. Therefore, we have that the first law of black hole mechanics~\cite{Bardeen:1973gs}
\begin{equation}
\delta\,\mathrm{M} = \mathrm{T}\,\delta\mathrm{S} \,, \label{first-law-general}
\end{equation}
is automatically satisfied if we vary with respect to the Komar mass $\mathrm{M}$, or, equivalently, if we identify the mass parameter $m$ with the actual mass by fixing the temporal gauge constant $\delta_{t}$ as
\begin{equation}
\delta_{t} = \frac{1}{\delta_\phi} \,. \label{first-law-New-NOT-accelerating-rn-br-condition}
\end{equation}
Moreover, this result is completely general, in the sense that in this derivation we have not used the condition for the removal of conical singularities~\eqref{New-semi-fixed-NOT-accelerating-rn-br-conical-condition}. Indeed, only the gauge constants $C_f$~\eqref{Ruffini-New-NOT-accelerating-rn-br-condition} and $\delta_t$~\eqref{first-law-New-NOT-accelerating-rn-br-condition} have been fixed, meaning that the azimuthal gauge parameter $\delta_\phi$ is a priori still free, and can still be used to remove the conical singularities as in equation~\eqref{New-semi-fixed-NOT-accelerating-rn-br-conical-condition}.

Additionally, it is interesting to point out that the result whereby the thermodynamic quantities are not affected by the presence of the Bertotti--Robinson field has also been found for the extremal Reissner--Nordstr\"{o}m black hole in Bertotti--Robinson in~\cite{DiPinto:2026rvp}, although in that case the black hole mass was determined using different methods.
\subsubsection{Summary and Combined Results for the New Regularized Metric}
\label{sec:summary-fixed-new-uncharged-non-accelerating-subcase}
Combining all the results of the previous sections, we obtain that the regularized form of the new parameterization is given by
\begin{subequations}
\label{New-fixed-NOT-accelerating-rn-br}
\begin{align}
{ds}^2 & = \frac{1}{\Omega^2}
\biggl[-\frac{\Delta_r}{r^2}\delta^2_{t}\,{dt}^2 + C^2_f\,r^2\biggl(\frac{{dr}^2}{\Delta_r}
+ \frac{{d\theta}^2}{\Delta_\theta}\biggr) + r^2\sin^2\theta\,\Delta_\theta\, \delta^2_{\phi}\,{d\phi}^2\biggr] \,, \label{New-fixed-NOT-accelerating-rn-br-metric} \\
A & = \frac{w}{B} \frac{\partial_\theta\Omega}{r\sin\theta}\,\delta_t\, dt
+ \frac{\sqrt{1-w^2}}{B} \bigl[\Omega - 1 - r\,\partial_r\Omega \bigr]\delta_{\phi}\, d\phi - \delta A_\phi\,\delta_{\phi}\, d\phi\,, \label{New-fixed-NOT-accelerating-rn-br-potential}
\end{align}
\end{subequations}
where
\begin{subequations}
\label{New-fixed-NOT-accelerating-rn-br-functions}
\begin{align}
\Delta_\theta & = 1 + B^2 m^2 \cos^2\theta \,, \\
\Delta_r & = \bigl[r^2(1-B^2 m^2)- 2 m r\bigr]\bigl[1+B^2(r-r_0)^2\bigr] \,, \\
\Omega^2 & = \bigl[1+ B^2(r-r_0)^2\bigr]- \bigl[r^2(1-B^2 m^2)- 2 m r\bigr]B^2\cos^2\theta \,, \\
\delta_{\phi} & = \frac{1}{\delta_{t}} = \frac{\sqrt{(1+B^2 m^2)^2-2 B^2 m (1-B^2 m^2)r_0}}{\sqrt{1+B^2 m^2}} \,,\\
C_f & = \sqrt{1+B^2 m^2}\sqrt{(1+B^2 m^2)^2-2 B^2 m (1-B^2 m^2)r_0} \,, \\
\delta A_\phi & =\sqrt{1-w^2} B\, m\, r_0 \,,
\end{align}
\end{subequations}
while the $r_0$ branches still correspond to
\begin{subequations}
\begin{align}
r_0 & = 0 \,, \label{r0-zero-not-charged-d} \tag{case I-d}\\
r_0 & = \frac{2 m}{1-B^2 m^2} \,. \label{r0-not-zero-not-charged-d} \tag{case II-d}
\end{align}
\end{subequations}
With the thermodynamical quantities resulting in
\begin{subequations}
\begin{align}
\mathrm{M} = m\,, \quad & \label{mass-New-fixed-NOT-accelerating-rn-br} \\
\kappa = \frac{1}{4m}\,,\quad &\Rightarrow \quad
\mathrm{T} = \frac{\kappa}{2\pi} = \frac{1}{8\pi m} \,, \label{surface-gravity-temperature-New-fixed-NOT-accelerating-rn-br} \\
\mathrm{A} = 16\pi m\,,\quad & \Rightarrow \quad\mathrm{S} = \frac{\mathrm{A}}{4} =4\pi m \,.\label{area-entropy-New-fixed-NOT-accelerating-rn-br}
\end{align}
\end{subequations}
In particular, as stated in the previous section, these quantities are the same as those for the Schwarzschild solution, independently of the Bertotti--Robinson parameter or the $r_0$ branch. Moreover, we also note that the various gauge parameters of the metric, namely $\delta_{t}$, $C_f$, and $\delta_{\phi}$, all reduce to $1$ in both the $B \to 0$ and $m \to 0$ limits, as expected.
\subsection{Geodesics and ISCO}
\label{sec:geodesics-isco-fixed-new-uncharged-non-accelerating-subcase}
Since the spacetime under consideration~\eqref{New-fixed-NOT-accelerating-rn-br} is stationary and axisymmetric, it follows that the Killing vectors $\xi=\partial_t$ and $\zeta=\partial_\varphi$ lead to the conservation of the energy $E$ and angular momentum $L$, respectively, of an uncharged test particle with four-momentum $p$:
\begin{equation}
\label{energy-angular-momentum-geodesic-general}
E = - g_{\mu \nu}\,p^{\mu}\,\xi^{\nu}\,, \quad \quad \quad L = g_{\mu \nu}\,p^{\mu}\,\zeta^{\nu}\,.
\end{equation}
Therefore, similarly to~\cite{Podolsky:2025tle,Astorino:2026okd}, we can study the geodesic motion in the equatorial plane, $\theta=\pi/2$, of a massive uncharged test particle with rest mass $m_0$ in the new regularized parameterization~\eqref{New-fixed-NOT-accelerating-rn-br}, for which the energy and angular momentum~\eqref{energy-angular-momentum-geodesic-general} are found to be
\begin{subequations}
\label{energy-angularmomentum-geodesic-New-fixed-NOT-accelerating-rn-br}
\begin{align}
\frac{E}{m_0}\bigg\rvert_{\theta=\frac{\pi}{2}} & = \biggl[(1-B^2 m^2)- \frac{2 m}{r}\biggr]\, \delta_t^2\,\dot{t} \label{energy-geodesic-New-fixed-NOT-accelerating-rn-br}\,, \\
\frac{L}{m_0}\bigg\rvert_{\theta=\frac{\pi}{2}} & = \frac{r^2}{1+ B^2(r-r_0)^2}\,\delta_\phi^2\,\dot{\phi} \,, \label{angular-momentum-geodesic-New-fixed-NOT-accelerating-rn-br}
\end{align}
\end{subequations}
where the dot denotes differentiation with respect to the affine parameter.

From the geodesic equation, $\nabla_u u=0$, and the metric compatibility, $\nabla g=0$, we obtain the additional conserved quantity $\chi=g_{\mu\nu}u^\mu u^\nu$, which takes the value $\chi=0$ for massless particles and $\chi=-1$ for massive ones.

Therefore, for a massive particle, using the computed expressions for the energy~\eqref{energy-geodesic-New-fixed-NOT-accelerating-rn-br} and angular momentum~\eqref{angular-momentum-geodesic-New-fixed-NOT-accelerating-rn-br}, the geodesic motion on the equatorial plane can be written as
\begin{equation}
\frac{1}{2}\,\dot{r}^2 = \frac{[1+B^2(r-r_0)^2]^2}{C_f^2\,\delta_t^2}\biggl[\frac{E}{2m_0}- V(r) \biggr] \,, \label{r-dot-New-fixed-NOT-accelerating-rn-br}
\end{equation}
where we introduced the effective potential $V(r)$ as
\begin{equation}
V(r) = \frac{1}{2}\biggl[(1-B^2 m^2)- \frac{2 m}{r}\biggr]\biggl[1+\frac{L^2\,[1+B^2(r-r_0)^2]}{m_0^2\,\delta_\phi^2\,r^2}\biggr]\,\delta_t^2 \,. \label{v-potential-New-fixed-NOT-accelerating-rn-br}
\end{equation}
Stable circular orbits are located at the minima of the potential $V(r)$~\eqref{v-potential-New-fixed-NOT-accelerating-rn-br}, while the \emph{innermost stable circular orbit} (ISCO) is defined by the conditions $V'(r)=V''(r)=0$, which, in our case, yield:
\begin{subequations}
\label{r-isco-L-isco-E-isco-New-fixed-NOT-accelerating-rn-br}
\begin{align}
r_\mathrm{ISCO} & = \frac{L^2_{\mathrm{ISCO}}\bigl[(1-B^2m^2)+6 m\,B^2\, r_0)\bigr]}{2m(m_0^2\delta_\phi^2+B^2L^2_{\mathrm{ISCO}})+2 B^2 r_0(1-B^2m^2) L^2_{\mathrm{ISCO}}} \,, \label{r-isco-New-fixed-NOT-accelerating-rn-br}\\
L_\mathrm{ISCO} & = \frac{2 \sqrt{3}(1+B^2 m\,r_0)\,m\,m_0\,\delta_{\phi}}{\sqrt{(1-B^2m^2)^2-12 B^2 m^2}} \,, \label{L-isco-New-fixed-NOT-accelerating-rn-br} \\
E_\mathrm{ISCO} & = \frac{2 \sqrt{2}(1-B^2 m^2)^\frac{3}{2}\,m_0\,\delta_{t}}{3(1+B^2m\,r_0)\sqrt{(1-B^2m^2)^2-12 B^2 m^2}} \,. \label{E-isco-New-fixed-NOT-accelerating-rn-br}
\end{align}
\end{subequations}
Moreover, by combining these results with the expression for the black hole horizon $r_H=\frac{2 m}{1-B^2 m^2}$~\eqref{horizon-New-NOT-accelerating-rn-br}, we find
\begin{equation}
\label{r-isco-B-New-fixed-NOT-accelerating-rn-br}
r_\mathrm{ISCO} = 3\biggl[\frac{1+B^2\,r_0^2}{1+3B^2\,r_0^2}\biggr]\biggl[\frac{2 m}{1-B^2m^2}\biggr] = 3\biggl[\frac{1+B^2\,r_0^2}{1+3B^2\,r_0^2}\biggr]r_H \,.
\end{equation}
Therefore, in the $r_0=0$ branch~\eqref{r0-zero-not-charged-d}, we obtain the same results as in~\cite{Podolsky:2025tle}, for which $r_\mathrm{ISCO} = 3\, r_H$, which is the same in form as the result for the Schwarzschild black hole. On the other hand, the general result~\eqref{r-isco-B-New-fixed-NOT-accelerating-rn-br} is different in the $r_0 \neq 0$ branch~\eqref{r0-not-zero-not-charged-d}, and indeed $r_\mathrm{ISCO} < 3\, r_H$ in the physical range of the parameters, namely $m>0$ and $|B|<\frac{1}{m}$.

Finally, for both $r_0$ branches, the addition of the Bertotti--Robinson electromagnetic field still increases the ISCO radius relative to the usual result for the asymptotically flat Schwarzschild solution, $r_\mathrm{ISCO} > r_{\mathrm{ISCO}-\mathrm{Schwarzschild}} = 6 m$, at least in the physical range of the parameters.
\section{Addition of the Melvin Parameter}
\label{sec:bonnor-melvin-addition}
With the new parameterization~\eqref{New-NOT-accelerating-rn-br} found in the previous section, we can add the Bonnor--Melvin external electromagnetic field to both physical $r_0$ branches simultaneously. For simplicity, we now consider the purely magnetic case, $w=0$, and add only the magnetic component of the Bonnor--Melvin field. The reason is that the purely electric counterpart can then be obtained by a simple duality rotation of the electromagnetic potential, leaving the metric unchanged. On the other hand, the electromagnetic generalization would not be static but stationary, due to the twist arising from the Lorentz force generated by the interaction between the electric (magnetic) component of the Bertotti--Robinson field and the magnetic (electric) component of the Bonnor--Melvin field~\cite{Astorino:2025lih}.
\subsection{The Ernst Method and the Harrison transformation}
\label{sec:ernst-harrison-theory}
The addition of the Melvin parameter can be obtained by means of the so-called Harrison transformation~\cite{Harrison:1968wue} within the Ernst method~\cite{Ernst:1967wx,Ernst:1967by}. The starting point of this method is the fact that, in the Einstein--Maxwell theory, a stationary and axisymmetric spacetime with a stationary and axisymmetric electromagnetic field can always be written, using cylindrical Weyl coordinates, in the Lewis--Weyl--Papapetrou (LWP) form\footnote{To be precise, there are two equivalent possible LWP forms, called the electric ansatz and the magnetic ansatz. The one given in equation~\eqref{magnetic-lwp-ansatz} is the magnetic one. Despite being equivalent, the application of the Ernst method leads to different results depending on the starting ansatz. In order to add the Melvin parameter, the seed solution must be written in the magnetic one. A modern comprehensive review of the Ernst method and the various transformations can be found in~\cite{Vigano:2022hrg,DiPinto:2024axv}.}:
\begin{subequations}
\label{magnetic-lwp-ansatz}
\begin{align}
{ds}^2 & = \frac{1}{f}\Bigl[-\rho^2 {dt}^2 + e^{2\gamma} \bigl({d\rho}^2 + {dz}^2\bigr)\Bigr] + f \bigl(d\phi - \omega\, d t\bigr)^2 \,, \label{magnetic-lwp-metric} \\
A & = A_t\, d t + A_\phi\, d\phi \,, \label{magnetic-lwp-potential}
\end{align}
\end{subequations}
from which one can define the gravitational twist potential $h$ and the electromagnetic twist potential $\tilde{A}_t$ by means of
\begin{subequations}
\label{twist-potentials}
\begin{align}
\label{Atilde}
\vec{e}_{\phi} \times  \vec{\nabla} \tilde{A}_{t} & =
- \frac{f}{\rho} \bigl( \vec{\nabla} A_{t} + \omega \vec{\nabla} A_{\phi} \bigr) \,, \\
\label{h}
\vec{e}_{\phi} \times \vec{\nabla} h & =
- \frac{f^2}{\rho} \vec{\nabla} \omega - 2\, \vec{e}_{\phi} \times \mathrm{Im} \bigl (\Phi^{*} \vec{\nabla}\Phi \bigr) \,,
\end{align}
\end{subequations}
and, finally, the complex gravitational Ernst potential $\pazocal{E}$ and the complex electromagnetic Ernst potential $\Phi$ as follows:
\begin{subequations}
\label{ernst-potentials}
\begin{align}
\Phi & = \tilde{A}_{t} - i A_{\phi} \,,\label{electromagnetic-ernst-potential} \\
\pazocal{E} & = - f - |\Phi|^2 + i h \,, \label{gravitational-ernst-potential}
\end{align}
\end{subequations}
where $\Vec{\nabla}$ and all the functions are understood with respect to the Euclidean space endowed with cylindrical coordinates $(\rho,\phi,z)$.

The introduction of the complex Ernst potentials~\eqref{ernst-potentials} allows the Einstein--Maxwell equations~\eqref{einstein-maxwell} to be recast as
\begin{subequations}
\label{ernst-em-equations}
\begin{align}
(\mathrm{Re}\,\pazocal{E} + |\Phi|^2) \, \nabla^2 \pazocal{E} & = \vec{\nabla} \pazocal{E} \cdot \bigl( \vec{\nabla} \pazocal{E} + 2 \Phi^* \vec{\nabla}\Phi \bigr) \,, \\
(\mathrm{Re}\,\pazocal{E} + |\Phi|^2) \, \nabla^2 \Phi & = \vec{\nabla} \Phi \cdot \bigl(\vec{\nabla} \pazocal{E} + 2 \Phi^* \vec{\nabla}\Phi \bigr) \,,
\end{align}
\end{subequations}
which enjoy a set of eight real Lie-point symmetries, corresponding to transformations of the complex potentials that leave the equations invariant. One of these symmetries is the Harrison transformation, which acts as
\begin{subequations}
\label{Harrison}
\begin{align}
\pazocal{E}' & = \frac{\pazocal{E}}{1 - 2\,\alpha^*\Phi - \alpha^* \alpha \, \pazocal{E}} \,,\, \label{Harrison-gravitationl} \\
\Phi' & = \frac{\alpha\,\pazocal{E} + \Phi}{1 - 2\,\alpha^*\Phi - \alpha^* \alpha \, \pazocal{E}} \,. \label{Harrison-electromagnetic}
\end{align}
\end{subequations}
When applied to a seed solution written in the magnetic ansatz~\eqref{magnetic-lwp-ansatz}, the Harrison transformation~\eqref{Harrison} adds the electromagnetic Melvin field to the solution. The strength of the added field is controlled by the complex parameter $\alpha$, with $\mathrm{Re}(\alpha)$ determining the electric component of the Melvin field and $\mathrm{Im}(\alpha)$ setting the magnetic one.
\subsection{Application of the Method to the New Parameterization}
\label{sec:ernst-harrison-application}
We now apply the Ernst method described above to the new parameterization of the uncharged and non-accelerating family of black holes in Bertotti--Robinson~\eqref{New-NOT-accelerating-rn-br}, restricting to the purely magnetic subcase $w=0$ and omitting the gauge constants $\delta_{t}$, $C_f$, $\delta_{\phi}$, and $\delta A_{\phi}$, which we will reintroduce and fix once again the new solution, generalized by the addition of the Melvin parameter, has been obtained.

We start by writing the seed solution~\eqref{New-NOT-accelerating-rn-br} in the LWP form~\eqref{magnetic-lwp-ansatz}, which can be achieved through the following transformation to cylindrical Weyl coordinates:
\begin{subequations}
\label{cylindrical-transformation-magnetic}
\begin{align}
\begin{split}
\rho & = \frac{\sin\theta\sqrt{\Delta_{r} \Delta_{\theta}}}{\Omega^2}  \\ 
& = \frac{\sin\theta \sqrt{\bigl[r^2(1-B^2 m^2)- 2 m r\bigr]\bigl[1+B^2(r-r_0)^2\bigr]\bigl[1 + B^2 m^2 \cos^2\theta\bigr]}}{\bigl[1+ B^2(r-r_0)^2\bigr]- \bigl[r^2(1-B^2 m^2)- 2 m r\bigr]B^2\cos^2\theta} \,,
\end{split}
\\
z & = \frac{\cos\theta\bigl[(r-m)(1+B^2 m\, r)-B^2 r_0\bigl[r^2(1-B^2 m^2)- 2 m r\bigr]-B^2\, m\, r_0^2\bigr]}{\bigl[1+ B^2(r-r_0)^2\bigr]- \bigl[r^2(1-B^2 m^2)- 2 m r\bigr]B^2\cos^2\theta} \,,
\end{align}
\end{subequations}
and by identifying the functions appearing in the LWP ansatz~\eqref{magnetic-lwp-ansatz} as
\begin{subequations}
\label{lwp-functions-seed-magnetic}
\begin{align}
f & = g_{\phi\phi} = \frac{r^2 \sin^2\theta \Delta_\theta}{\Omega^2}\,, \label{seed-f}\\
\omega & = A_{t}= 0\,, \label{seed-omega-At}\\
A_{\phi}& =\frac{1}{B} \bigl[\Omega - 1 - r\,\partial_r\Omega \bigr] \label{seed-Aphis}\,,
\end{align}
\end{subequations}
where the functions $\Delta_\theta$, $\Delta_r$, and $\Omega$ are still those given by equation~\eqref{New-NOT-accelerating-rn-br-functions}.

Since the spacetime under consideration is purely magnetic, $A_t=0$, and non-rotating, $\omega = 0$, the twist potentials~\eqref{twist-potentials} reduce to gauge constants, which can be set to zero,
\begin{equation}
\tilde{A}_t = h = 0 \,. \label{twist-potentials-seed-magnetic}
\end{equation}
The complex Ernst potentials~\eqref{ernst-potentials} in this case then take the form
\begin{subequations}
\label{ernst-potentials-seed-magnetic}
\begin{align}
\Phi & = - i A_{\phi} \,,\label{electromagnetic-ernst-potential-seed-magnetic} \\
\pazocal{E} & = \frac{2i}{B} \Phi - r_0^2 = \frac{2}{B}A_{\phi}-r_0^2 \,, \label{gravitational-ernst-potential-seed-magnetic}
\end{align}
\end{subequations}
which, after a Harrison transformation~\eqref{Harrison} with imaginary parameter $\alpha = i\, b$, become
\begin{subequations}
\label{Ernst-potentials-harrison-magnetic}
\begin{align}
\pazocal{E}' & = \frac{2A_{\phi}-B\,r_0^2}{B + 2b\,(B-b) A_{\phi} + b^2\,B\,r_0^2} \,,\, \label{Harrison-gravitational-ernst-potential} \\
\Phi' & = -i\biggl[\frac{(B-2 b)A_{\phi} + b\,B\, r_0^2}{B + 2b\,(B-b) A_{\phi} + b^2\,B\,r_0^2}\biggr] \,. \label{Harrison-electromagnetic-ernst-potential}
\end{align}
\end{subequations}
As the Harrison-transformed gravitational Ernst potential~\eqref{Harrison-gravitational-ernst-potential} is purely real, whereas the electromagnetic Ernst potential is purely imaginary, the definition of the complex Ernst potentials~\eqref{ernst-potentials} implies that the new Harrison-transformed twist potentials~\eqref{twist-potentials} remain zero:
\begin{equation}
\tilde{A}'_t = h' = 0 \,, \label{twist-potentials-Harrison-magnetic}
\end{equation}
while the remaining functions of the LWP ansatz~\eqref{magnetic-lwp-ansatz} are therefore given by
\begin{subequations}
\label{lwp-functions-Harrison-magnetic}
\begin{align}
f' & = \frac{ r^2 \sin^2\theta \Delta_\theta}{g^2} \,,\\
A'_{\phi} & = \frac{(B-2b)\bigl[\Gamma-\Omega\bigr]+ (b\, B^2 r_0^2) \Omega}{B^2 g} \\
A_t' & = \omega' = 0 \,,
\end{align}
\end{subequations}
where the new functions $g$ and $\Gamma$ correspond to
\begin{subequations}
\label{s-br-bm-new-functions}
\begin{align}
g & = \frac{1}{B^2}\Bigl[2 b (B - b)\Gamma  + \bigr[(2 b^2 - 2 b B + B^2) + b^2 B^2 r_0^2\bigl]\Omega \Bigr] \,, \\
\Gamma & = 1 + B^2 \bigl[m r \cos^2\theta  - r_0(r-r_0)\bigr] \,,
\end{align}
\end{subequations}
and the other functions retain the form given in equations~\eqref{New-NOT-accelerating-rn-br-functions}.
\subsection{The New Solution in a Bertotti--Robinson--Bonnor--Melvin Universe}
\label{sec:new-uncharged-non-accelerating-subcase-with-Melvin}
As a summary, we collect here the metric and electromagnetic potential resulting from the computations presented above, which thus define a two-branch family of solutions, both describing a massive Schwarzschild black hole immersed in a combination of Bertotti--Robinson and Bonnor--Melvin magnetic fields:
\begin{subequations}
\label{New-s-br-bm}
\begin{align}
{ds}^2 & = -\frac{g^2\Delta_r}{r^2 \Omega^2}{dt}^2 + \frac{g^2 r^2}{\Omega^2}\biggl[\frac{{d\theta}^2}{\Delta_\theta}+\frac{{dr}^2}{\Delta_r}\biggl]+ \frac{ r^2 \sin^2\theta \Delta_\theta}{g^2}{d\phi}^2 \,, \label{New-s-br-bm-metric}\\
A & = \biggl[\frac{(B-2b)\bigl[\Gamma-\Omega\bigr]+ (b\, B^2 r_0^2) \Omega}{B^2 g}\biggl]d\phi \,, \label{New-s-br-bm-potential}
\end{align}
\end{subequations}
where
\begin{subequations}
\label{New-s-br-bm-functions}
\begin{align}
\Delta_\theta & = 1 + B^2 m^2 \cos^2\theta \,, \\
\Delta_r & = \bigl[r^2(1-B^2 m^2)- 2 m r\bigr]\bigl[1+B^2(r-r_0)^2\bigr] \,, \\
\Omega^2 & = \bigl[1+ B^2(r-r_0)^2\bigr]- \bigl[r^2(1-B^2 m^2)- 2 m r\bigr]B^2\cos^2\theta \,, \\
g & = \frac{1}{B^2}\Bigl[2 b (B - b)\Gamma  \,+\, \bigr[(2 b^2 - 2 b B + B^2) + b^2 B^2 r_0^2\bigl]\Omega \Bigr] \,, \\
\Gamma & = 1 + B^2 \bigl[m r \cos^2\theta  - r_0(r-r_0)\bigr] \,,
\end{align}
\end{subequations}
with the two possible $r_0$ branches still given by
\begin{align}
r_0 & = 0 \,, \label{r0-zero-not-charged-e} \tag{case I-e}\\
r_0 & = \frac{2 m}{1-B^2 m^2} \,, \label{r0-not-zero-not-charged-e} \tag{case II-e}
\end{align}
and where $m$ is the mass parameter, while $b$ and $B$ determine the values of the Bonnor--Melvin and Bertotti--Robinson fields, respectively.

In the $r_0=0$ branch~\eqref{r0-zero-not-charged-d}, this solution reduces to the one found by Astorino in~\cite{Astorino:2025lih}, while the $r_0\neq0$ branch~\eqref{r0-not-zero-not-charged-d} constitutes a genuinely new family of solutions to the Einstein--Maxwell equations~\eqref{einstein-maxwell}.

Nonetheless, we stress that only thanks to the new parameterization~\eqref{New-NOT-accelerating-rn-br} have we been able to add the Melvin parameter $b$ to both physical branches simultaneously. Had we instead started from the non-accelerating subcase of the static Ovcharenko--Podolsk\'{y} parameterization~\eqref{NOT-accelerating-rn-br}, we would have obtained an unphysical $r_0\neq0$ branch corresponding to the Melvin generalization of the $\mathrm{RN}-\mathrm{BR}_0$ solution, as discussed in section~\ref{sec:uncharged-non-accelerating-subcase}.

Finally, since this solution generalizes the Schwarzschild black hole in the Melvin universe, we can immediately conclude that it is of Petrov type $I$.
\subsection{Regularized Form of the New Solution}
\label{sec:fixed-new-uncharged-non-accelerating-subcase-with-Melvin}
Having obtained the new solution~\eqref{New-s-br-bm} with both the Bertotti--Robinson and Bonnor--Melvin magnetic fields, we can now proceed as in section~\eqref{sec:fixed-new-uncharged-non-accelerating-subcase} by introducing the gauge constants $\delta_t$, $C_f$, $\delta_\phi$, and $\delta A_\phi$, to remove any possible pathologies, if necessary, as follows:
\begin{subequations}
\label{semi-fixed-New-s-br-bm}
\begin{align}
{ds}^2 & = -\frac{g^2\Delta_r}{r^2 \Omega^2}\delta^2_{t}\,{dt}^2 + \frac{C_f^2\,g^2 r^2}{\Omega^2}\biggl[\frac{{d\theta}^2}{\Delta_\theta}+\frac{{dr}^2}{\Delta_r}\biggl]+ \frac{ r^2 \sin^2\theta \Delta_\theta}{g^2}\delta^2_{\phi}\,{d\phi}^2 \,, \label{semi-fixed-New-s-br-bm-metric}\\
A & = \biggl[\frac{(B-2b)\bigl[\Gamma-\Omega\bigr]+ (b\, B^2 r_0^2) \Omega}{B^2 g}-\delta A_{\phi}\biggl]\delta_{\phi}\,d\phi \,, \label{semi-fixed-New-s-br-bm-potential}
\end{align}
\end{subequations}
with the various remaining functions still as given in equation~\eqref{New-s-br-bm-functions}.

In this ansatz, we find that it is possible to remove the conical singularities~\eqref{conical-condition-0} if
\begin{equation}
\label{New-semi-fixed-s-br-bm-conical-condition}
\delta_{\phi} = \frac{C_f (1+ b\, B\, m\, r_0)^4}{(1+B^2m^2)}  \,.
\end{equation}
Similarly, the Dirac strings~\eqref{dirac-condition} can also be removed by setting
\begin{equation}
\label{New-semi-fixed-s-br-bm-dirac-condition}
\delta A_\phi = \frac{B\, m\, r_0}{(1+ b\, B\, m\, r_0)}\,,
\end{equation}
while, by the same arguments as in the non-Melvin subcase, this solution does not exhibit Misner strings~\eqref{misner-condition} or closed timelike curves~\eqref{ctcs-condition}.

Regarding the thermodynamics, all the results found in section~\ref{sec:fixed-new-uncharged-non-accelerating-subcase-thermodynamics} remain unaffected by the addition of the Melvin field, so that the Smarr law~\eqref{smarr-law-New-NOT-accelerating-rn-br} is automatically satisfied, the Christodoulou--Ruffini mass formula~\eqref{Ruffini-general} holds upon imposing the same condition, namely \begin{equation}
C_f = \bigl[(1+B^2 m^2)^2-2 B^2 m (1-B^2 m^2)r_0\bigr]\delta^2_{t} \delta_{\phi} \label{Ruffini-New-s-br-bm-condition}\,,
\end{equation}
while the first law of black hole mechanics~\eqref{first-law-general} is satisfied with respect to the mass parameter $m$ by fixing the temporal gauge constant $\delta_t$ as
\begin{equation}
\delta_{t} = \frac{1}{\delta_\phi} \,. \label{first-law-New-s-br-bm-condition}
\end{equation}
\subsubsection{Summary and Combined Results for the New Regularized Solution}
\label{sec:summary-fixed-new-uncharged-non-accelerating-subcase-with-Melvin}
Therefore, combining the results obtained above, we can finally write explicitly the regularized form of the new solution~\eqref{New-s-br-bm}, which describes a two-branch family of uncharged and non-accelerating black hole solutions with an external Bertotti--Robinson--Bonnor--Melvin magnetic field:
\begin{subequations}
\label{fixed-New-s-br-bm}
\begin{align}
{ds}^2 & = -\frac{g^2\Delta_r}{r^2 \Omega^2}\delta^2_{t}\,{dt}^2 + \frac{C_f^2\,g^2 r^2}{\Omega^2}\biggl[\frac{{d\theta}^2}{\Delta_\theta}+\frac{{dr}^2}{\Delta_r}\biggl]+ \frac{ r^2 \sin^2\theta \Delta_\theta}{g^2}\delta^2_{\phi}\,{d\phi}^2 \,, \label{fixed-New-s-br-bm-metric}\\
A & = \biggl[\frac{(B-2b)\bigl[\Gamma-\Omega\bigr]+ b\, B^2 r_0^2\, \Omega}{B^2 g}-\delta A_{\phi}\biggl]\delta_{\phi}\,d\phi \,, \label{fixed-New-s-br-bm-potential}
\end{align}
\end{subequations}
where
\begin{subequations}
\label{fixed-New-s-br-bm-functions}
\begin{align}
\Delta_\theta & = 1 + B^2 m^2 \cos^2\theta \,, \\
\Delta_r & = \bigl[r^2(1-B^2 m^2)- 2 m r\bigr]\bigl[1+B^2(r-r_0)^2\bigr] \,, \\
\Omega^2 & = \bigl[1+ B^2(r-r_0)^2\bigr]- \bigl[r^2(1-B^2 m^2)- 2 m r\bigr]B^2\cos^2\theta \,, \\
g & = \frac{1}{B^2}\Bigl[2 b (B - b)\Gamma  + \bigr[(2 b^2 - 2 b B + B^2) + b^2 B^2 r_0^2\bigl]\Omega \Bigr] \,, \\
\Gamma & = 1 + B^2 \bigl[m r \cos^2\theta  - r_0(r-r_0)\bigr] \,, \\
\delta_{\phi} & = \frac{1}{\delta_{t}} = \frac{(1+ b\, B\, m\, r_0)^2}{\sqrt{1+B^2 m^2}}\sqrt{(1+B^2 m^2)^2-2 B^2 m (1-B^2 m^2)r_0} \,,\\
C_f & = \frac{\sqrt{1+B^2 m^2}}{(1+ b\, B\, m\, r_0)^2}\sqrt{(1+B^2 m^2)^2-2 B^2 m (1-B^2 m^2)r_0} \,,\\
\delta A_\phi & =\frac{B\, m\, r_0}{(1+ b\, B\, m\, r_0)} \,,
\end{align}
\end{subequations}
with
\begin{subequations}
\begin{align}
r_0 & = 0 \,, \label{r0-zero-not-charged-f} \tag{case I-f}\\
r_0 & = \frac{2 m}{1-B^2 m^2} \,. \label{r0-not-zero-not-charged-f} \tag{case II-f}
\end{align}
\end{subequations}
Moreover, the charges and thermodynamical quantities remain those of the Schwarzschild solution without any external electromagnetic field:
\begin{subequations}
\begin{align}
\mathrm{Q} = \mathrm{P} = 0 \,, \quad & \label{charges-s-br-bm}\\
\mathrm{M} = m\,, \quad & \label{mass-s-br-bm} \\
\kappa = \frac{1}{4m}\,,\quad &\Rightarrow \quad
\mathrm{T} = \frac{\kappa}{2\pi} = \frac{1}{8\pi m} \,, \label{surface-gravity-temperature-New-fixed-s-br-bm} \\
\mathrm{A} = 16\pi m\,,\quad & \Rightarrow \quad\mathrm{S} = \frac{\mathrm{A}}{4} = 4\pi m \,.\label{area-entropy-New-fixed-s-br-bm}
\end{align}
\end{subequations}
\section{Vacuum Subcase(s) of the Black Hole in Bertotti--Robinson--Bonnor--Melvin}
\label{sec:vacuum-subcase-s-br-bm}
As found by Astorino in~\cite{Astorino:2025lih,Astorino:2026okd}, it is possible to tune the Melvin parameter $b$ in the $r_0=0$ branch~\eqref{r0-zero-not-charged-f} of the solution which contains both the Bertotti--Robinson and Bonnor--Melvin fields~\eqref{New-s-br-bm} in such a way that the electromagnetic field vanishes, $A=0$, thereby yielding a vacuum solution of the Einstein equations, $R_{\mu \nu}=0$, which corresponds to a non-asymptotically flat, Petrov type $I$ generalization of the Schwarzschild black hole, now embedded in an external gravitational field. Moreover, an additional solution can also be obtained by the analytic continuation $B \mapsto i B$ of this tuned vacuum solution, which provides a representation in spherical coordinates of the Schwarzschild black hole inside the expanding bubble of nothing~\cite{Witten:1981gj,Aharony:2002cx,Astorino:2022fge,Astorino:2025tqg}.

We therefore present here the resulting metric and the corresponding value of the Melvin parameter that allows the same procedure to be performed simultaneously for both $r_0$ branches, and in particular for the $r_0\neq0$ branch~\eqref{r0-not-zero-not-charged-f}.

However, as we will prove, the parameter $r_0$ appearing in the resulting metrics will actually be inessential, as it can be removed via a coordinate transformation. Nonetheless, the $r_0\neq 0$ branch can still be regarded as an alternative parameterization of these vacuum solutions in spherical coordinates, which may be more useful for certain computations than its $r_0=0$ counterpart.

At first sight, it may not be straightforward that there exists a value of the Melvin parameter $b$ for which the electromagnetic potential $A$~\eqref{fixed-New-s-br-bm-potential} vanishes. Nevertheless, by substituting the value of the gauge parameter $\delta A_\phi$ that removes the Dirac strings~\eqref{New-semi-fixed-s-br-bm-dirac-condition} and performing some algebraic manipulations, one finds that the potential can be rewritten as
\begin{equation}
A = \biggl[\frac{B}{2+B^2m\,r_0}-b\biggr]\biggl[\frac{(2+B^2m\,r_0)\bigl[\Gamma-\Omega\bigr]- B^2 r_0^2\, \Omega}{(1+b\,B\,m\,r_0)B^2 g}\biggl]\delta_{\phi}\,d\phi \,. \label{fixed-New-s-br-bm-potential-pre-vacuum}
\end{equation}
Therefore, one obtains a vacuum solution, $A=0$, by setting
\begin{equation}
b = \frac{B}{2+B^2m\,r_0}\,. \label{fixed-New-s-br-bm-vacuum-condition}
\end{equation}

\subsection{The Schwarzschild Black Hole Immersed in an External Gravitational Field}
\label{sec:s-br-bm-vacuum-egf}
The resulting spacetime obtained by tuning the Melvin parameter as in equation~\eqref{fixed-New-s-br-bm-vacuum-condition} is given by
\begin{subequations}
\label{fixed-New-s-egf-vacuum}
\begin{align}
{ds}^2 & = -\frac{g^2\Delta_r}{r^2 \Omega^2}\delta^2_{t}\,{dt}^2 + \frac{C_f^2\,g^2 r^2}{\Omega^2}\biggl[\frac{{d\theta}^2}{\Delta_\theta}+\frac{{dr}^2}{\Delta_r}\biggl]+ \frac{ r^2 \sin^2\theta \Delta_\theta}{g^2}\delta^2_{\phi}\,{d\phi}^2 \,, \label{fixed-New-s-egf-vacuum-metric}\\
\Delta_\theta & = 1 + B^2 m^2 \cos^2\theta \,, \\
\Delta_r & = \bigl[r^2(1-B^2 m^2)- 2 m r\bigr]\bigl[1+B^2(r-r_0)^2\bigr] \,, \\
\Omega^2 & = \bigl[1+ B^2(r-r_0)^2\bigr]- \bigl[r^2(1-B^2 m^2)- 2 m r\bigr]B^2\cos^2\theta \,, \\
g & = \frac{2(1+B^2\,m\,r_0)}{(2+B^2\,m\,r_0)^2}\bigl[\Gamma +(1+B^2\,m\,r_0)\Omega\bigr] \,, \\
\Gamma & = 1 + B^2 \bigl[m r \cos^2\theta  - r_0(r-r_0)\bigr] \,, \\
\delta_{\phi} & = \frac{1}{\delta_{t}} = \frac{4(1+ B^2\,m\,r_0)^2}{(2+ B^2\,m\,r_0)^2\sqrt{1+B^2 m^2}}\sqrt{(1+B^2 m^2)^2-2 B^2 m (1-B^2 m^2)r_0} \,,\\
C_f & = \frac{(2+ B^2\,m\,r_0)^2\sqrt{1+B^2 m^2}}{4(1+ B^2\,m\,r_0)^2}\sqrt{(1+B^2 m^2)^2-2 B^2 m (1-B^2 m^2)r_0} \,,
\end{align}
\end{subequations}
with the two $r_0$ branches still corresponding to:
\begin{subequations}
\begin{align}
r_0 & = 0 \,, \label{r0-zero-not-charged-g} \tag{case I-g}\\
r_0 & = \frac{2 m}{1-B^2 m^2} \,. \label{r0-not-zero-not-charged-g} \tag{case II-g}
\end{align}
\end{subequations}
This vacuum subcase describes a Schwarzschild black hole of radius $r_H=\frac{2m}{1-B^2m^2}$ immersed in an external gravitational field. Indeed, for both $r_0$ branches, it reduces to the Schwarzschild solution~\eqref{positive-Schwarzschild} in the $B\to0$ limit, while for $m\to0$ it reduces to a non-asymptotically flat purely gravitational background of Petrov type $D$, determined solely by the constant $B$, whose metric we report here for completeness:
\begin{subequations}
\label{br-bm-egt-vacuum}
\begin{align}
{ds}^2 & = -\frac{g^2\Delta_r}{r^2 \Omega^2}{dt}^2 + \frac{g^2 r^2}{\Omega^2}\biggl[{d\theta}^2+\frac{{dr}^2}{\Delta_r}\biggl]+ \frac{ r^2 \sin^2\theta }{g^2}\,{d\phi}^2 \,, \label{br-bm-egt-metric}\\
\Delta_r & = r^2\bigl[1+B^2r^2\bigr] \,, \\
\Omega^2 & = 1+ B^2r^2\sin^2\theta \,, \\
g & = \frac{1}{2}(1 +\Omega) \,.
\end{align}
\end{subequations}
\subsection{The Schwarzschild Black Hole Inside the Expanding Bubble of Nothing}
\label{sec:s-br-bm-vacuum-bubble}
As the vacuum black hole solution written in the previous section~\eqref{fixed-New-s-egf-vacuum} depends only on even powers of the parameter $B$, it follows that another solution of the vacuum Einstein equations can be obtained by the analytic continuation $B \mapsto i B$. Since this solution will be used in the next section, we report the explicit result here:
\begin{subequations}
\label{fixed-New-s-bubble-vacuum}
\begin{align}
{ds}^2 & = -\frac{g^2\Delta_r}{r^2 \Omega^2}\delta^2_{t}\,{dt}^2 + \frac{C_f^2\,g^2 r^2}{\Omega^2}\biggl[\frac{{d\theta}^2}{\Delta_\theta}+\frac{{dr}^2}{\Delta_r}\biggl]+ \frac{ r^2 \sin^2\theta \Delta_\theta}{g^2}\delta^2_{\phi}\,{d\phi}^2 \,, \label{fixed-New-s-bubble-vacuum-metric}\\
\Delta_\theta & = 1 - B^2 m^2 \cos^2\theta \,, \\
\Delta_r & = \bigl[r^2(1+B^2 m^2)- 2 m r\bigr]\bigl[1-B^2(r-r_0)^2\bigr] \,, \\
\Omega^2 & = \bigl[1- B^2(r-r_0)^2\bigr]+ \bigl[r^2(1+B^2 m^2)- 2 m r\bigr]B^2\cos^2\theta \,, \\
g & = \frac{2(1-B^2\,m\,r_0)}{(2-B^2\,m\,r_0)^2}\bigl[\Gamma +(1-B^2\,m\,r_0)\Omega\bigr] \,, \\
\Gamma & = 1 - B^2 \bigl[m r \cos^2\theta  - r_0(r-r_0)\bigr] \,, \\
\delta_{\phi} & = \frac{1}{\delta_{t}} = \frac{4(1- B^2\,m\,r_0)^2}{(2- B^2\,m\,r_0)^2\sqrt{1-B^2 m^2}}\sqrt{(1-B^2 m^2)^2+2 B^2 m (1+B^2 m^2)r_0} \,,\\
C_f & = \frac{(2- B^2\,m\,r_0)^2\sqrt{1-B^2 m^2}}{4(1- B^2\,m\,r_0)^2}\sqrt{(1-B^2 m^2)^2+2 B^2 m (1+B^2 m^2)r_0} \,,
\end{align}
\end{subequations}
where the two $r_0$ branches are now given by:
\begin{subequations}
\begin{align}
r_0 & = 0 \,, \label{r0-zero-not-charged-h} \tag{case I-h}\\
r_0 & = \frac{2 m}{1+B^2 m^2} \,. \label{r0-not-zero-not-charged-h} \tag{case II-h}
\end{align}
\end{subequations}
This metric is the expression in spherical coordinates for the Schwarzschild black hole, now with radius $r_H=\frac{2m}{1+B^2m^2}$, inside the expanding bubble of nothing, first discovered in~\cite{Astorino:2022fge}. Indeed, it reduces to the Schwarzschild solution~\eqref{positive-Schwarzschild} in the $B \to 0$ limit, while for $m \to 0$ it becomes
\begin{subequations}
\label{br-bm-bubble-vacuum}
\begin{align}
{ds}^2 & = -\frac{g^2\Delta_r}{r^2 \Omega^2}{dt}^2 + \frac{g^2 r^2}{\Omega^2}\biggl[{d\theta}^2+\frac{{dr}^2}{\Delta_r}\biggl]+ \frac{ r^2 \sin^2\theta }{g^2}\,{d\phi}^2 \,, \label{br-bm-bubble-metric}\\
\Delta_r & = r^2\bigl[1-B^2r^2\bigr] \,, \\
\Omega^2 & = 1 - B^2r^2\sin^2\theta \,, \\
g & = \frac{1}{2}(1 +\Omega) \,,
\end{align}
\end{subequations}
that, as shown in~\cite{Astorino:2026okd}, can be transformed by means of the following coordinate transformation:
\begin{subequations}
\label{transf-bubble}
\begin{align}
r & \mapsto \frac{2m\sqrt{\tilde{r}^2-2m\tilde{r}+m^2\cos^2\vartheta}}{\tilde{r}-m}\,,\\
\theta & \mapsto \arccos\biggl(\frac{m\cos\vartheta}{\sqrt{\tilde{r}^2-2m\tilde{r}+m^2\cos^2\vartheta}}\biggr) \,,\\
t & \mapsto 2 m \tau \,, \quad \phi \mapsto \frac{\phi}{4 m}\,, \quad B \mapsto \frac{1}{2m} \,,
\end{align}
\end{subequations}
into the double--Wick rotation ($t \mapsto i \varphi$, $\phi \mapsto i \tau$) of the Schwarzschild solution, which is precisely the metric describing an expanding bubble of nothing~\cite{Witten:1981gj,Aharony:2002cx}.
\begin{equation}
\label{positive-bubble}
{ds}^2 = -\tilde{r}^2\sin^2\vartheta\, {d\tau}^2 + \frac{{d\tilde{r}}^2}{\bigl(1-\frac{2m}{\tilde{r}}\bigr)}
+ \tilde{r}^2{d\vartheta}^2 + \biggl(1-\frac{2m}{\tilde{r}}\biggr){d\varphi}^2 \,.
\end{equation}
A more rigorous proof that this vacuum subcase~\eqref{fixed-New-s-bubble-vacuum} can be mapped to the black hole inside the expanding bubble of nothing was first given in~\cite{Astorino:2025tqg} for the $r_0=0$ branch, while we will prove it for the $r_0 \neq 0$ branch in the next section.
\subsection{Inessentiality of the \texorpdfstring{$r_0$}{r0} Parameter for the Vacuum Subcase}
\label{sec:s-br-bm-r0-inessential}
The metric describing the Schwarzschild black hole inside the expanding bubble of nothing~\cite{Astorino:2022fge} can be written in cylindrical Weyl coordinates $(t,\rho,\phi,z)$ by means of the basic solitonic bricks $\mu_i(\rho,z)$ as:
\begin{equation}
\label{s-bubble-weyl}
{ds}^2 = - \rho^2\frac{\mu_2\, \mu_4}{\mu_1\,\mu_3}\,\tilde{\delta}_{t}^2\, {dt}^2 + \frac{\tilde{C}^2_f \,  \mu_1^3\, \mu_2^5\, \mu_3^3\, \mu_4^5({d\rho}^2+{dz}^2)}{ \mu_{12}\, \mu_{14}\, \mu_{23}\, \mu_{34}\, W_{11}\, W_{22}\, W_{33}\, W_{44}\, W_{13}^2\, W_{24}^2} \, + \frac{\mu_1\,\mu_3}{\mu_2\, \mu_4}\,\tilde{\delta}_{\phi}^2\, {d\phi}^2 \,,
\end{equation}
with
\begin{subequations}
\label{solitons}
\begin{align}
\mu_i & = w_i-z+\sqrt{\rho^2 + (z-w_i)^2} \,, \\
\mu_{ij} & = (\mu_{i}-\mu_{j})^2  \,, \\
W_{ij} & = \rho^2 + \mu_i\,\mu_j \,,
\end{align}
\end{subequations}
where $w_2$ and $w_3$ represent the poles of the Schwarzschild black hole, $w_1$ and $w_4$ are those of the bubble of nothing, while $\tilde{\delta}_t$, $\tilde{C}_f$ and $\tilde{\delta}_\phi$ are gauge constants.

We find that the vacuum subcase obtained by the analytic continuation $B \mapsto iB$, corresponding to equations~\eqref{fixed-New-s-bubble-vacuum}, can be mapped to the metric in Weyl coordinates given above~\eqref{s-bubble-weyl} by means of the following coordinate transformation:
\begin{subequations}
\begin{align}
\rho & = \frac{\sin\theta \sqrt{\bigl[r^2(1+B^2 m^2)- 2 m r\bigr]\bigl[1-B^2(r-r_0)^2\bigr]\bigl[1 - B^2 m^2 \cos^2\theta\bigr]}}{\bigl[1- B^2(r-r_0)^2\bigr]+ \bigl[r^2(1+B^2 m^2)- 2 m r\bigr]B^2\cos^2\theta} \,, \\
z & = \frac{\cos\theta\bigl[(r-m)(1-B^2 m\, r)+B^2 r_0\bigl[r^2(1+B^2 m^2)- 2 m r\bigr]+B^2\, m\, r_0^2\bigr]}{\bigl[1- B^2(r-r_0)^2\bigr]+ \bigl[r^2(1+B^2 m^2)- 2 m r\bigr]B^2\cos^2\theta} \,,
\end{align}
\end{subequations}
and by identifying the gauge constants as:
\begin{subequations}
\label{gauge-const-bubble}
\begin{align}
\tilde{\delta}_\phi & = \frac{1}{\tilde{\delta}_t} =\biggl[\frac{(2-B^2\,m\,r_0)^2}{2 B(1-B^2\,m\,r_0)}\biggr]{\delta}_\phi\,,\\
\tilde{C}_f & = \frac{128\,C_f\, m(1-B^2\,m\,r_0)^3}{B^3(2-B^2\,m\,r_0)^2}\,,
\end{align}
\end{subequations}
and the poles $w_i$ as
\begin{equation}
\label{poles-bubble}
 w_1 = - \frac{1}{B} \,, \qquad w_2 = - m  \,, \qquad w_3 = m \,, \qquad w_4 = \frac{1}{B} \,.
\end{equation}
Therefore, we find that \emph{both} $r_0$ branches of the analytic continued vacuum subcase~\eqref{fixed-New-s-bubble-vacuum} are actually mapped to the \emph{same} metric in Weyl coordinates. In other words, the two branches are diffeomorphic, since the $r_0$ parameter can be introduced or removed by a transformation in Weyl coordinates, followed by the inverse transformation mapping to the other $r_0$ branch. 

Obviously, by analytic continuation, the same argument also applies to the metric describing the Schwarzschild black hole immersed in the external gravitational field~\eqref{fixed-New-s-egf-vacuum}.

The fact that the two $r_0$ branches actually correspond to the same solution in the vacuum subcase may seem surprising at first. Physically, however, this can be understood by recalling, as explained in~\cite{Ovcharenko:2026byw} and at the end of section~\ref{sec:charged-static-family}, that the parameter $r_0$ controls whether the aligned component of the electromagnetic field is determined by the non-aligned one or is instead independent of it. Therefore, since we are considering the vacuum subcases, i.e.~\emph{without} an electromagnetic field, it is only natural that the $r_0$ parameter becomes inessential in these subcases.

\section{Conclusions}
\label{sec:conclusions}
In this work, we have presented a new parameterization for the subfamily of exact Einstein--Maxwell solutions found by Ovcharenko--Podolsk\'{y} that describes uncharged and non-accelerating static black holes immersed in an external Bertotti--Robinson electromagnetic field. This new parameterization of the family is, in general, different from the original one, except for the $r_0=0$ subcase, for which both families reduce to the Schwarzschild$-\mathrm{BR}$ solution. In particular, while in the original parameterization the $r_0 \neq 0$ branch solving the field equations, namely the $\mathrm{RN}-\mathrm{BR}_0$ solution, contains subcases describing black holes with a \emph{negative} mass, the new parameterization always yields black holes with a \emph{positive} mass on both $r_0$ branches.

Subsequently, we have introduced additional gauge constants and fixed them appropriately. Some of these have been fixed so that the spacetime can be made free of conical singularities and Dirac strings, while Misner strings and closed timelike curves were not already present. The remaining constants have instead been fixed in such a way as to satisfy the first law of black hole mechanics, the Smarr law, and the Christodoulou--Ruffini mass formula, but only if the black hole mass is computed not using the usual Komar prescription, which leads to ill-defined unphysical results in this case due to the fact that the spacetime is not asymptotically flat, but as a Komar integral evaluated on the black hole horizon. Quite surprisingly, in this way we have found that the mass and all the thermodynamic quantities exactly coincide with the corresponding Schwarzschild results and are not modified by the presence of the Bertotti--Robinson field. 

Moreover, we have studied the geodesic motion of massive uncharged test particles in the equatorial plane, determining the analytic form of the innermost stable circular orbit (ISCO). In particular, for the $r_0=0$ branch, we have recovered the already known result that the ISCO radius is three times the radius of the black hole horizon, while for the $r_0\neq0$ branch we have found a different result, namely that the ISCO radius is always smaller than the corresponding $r_0=0$ value, at least in the physical range of the parameters. Nevertheless, for both $r_0$ branches and within the same physical range of the parameters, the addition of the Bertotti--Robinson electromagnetic field increases both the ISCO radius and the black hole horizon radius when compared with the Schwarzschild black hole without any external electromagnetic field.

We then generalized the new parameterization, in the purely magnetic subcase, by also introducing the Melvin parameter through a Harrison transformation within the Ernst method. The resulting solution thus describes a massive, uncharged, and non-accelerating static black hole immersed in a Bertotti--Robinson--Bonnor--Melvin universe. In particular, the $r_0=0$ branch of this new family corresponds to the one recently found by Astorino, while the $r_0\neq0$ branch is a genuinely new solution of the Einstein--Maxwell equations. Additionally, as in the previous purely Bertotti--Robinson case, we introduced and fixed appropriate gauge constants in order to remove all possible pathologies and satisfy the laws of black hole thermodynamics. In particular, we have found that the addition of the Melvin parameter also does \emph{not} modify the charges or the thermodynamic quantities, which therefore remain those of the Schwarzschild black hole.

Furthermore, we have also investigated the vacuum subcase of the new family, which, in accordance with the results already found by Astorino for the $r_0=0$ branch, represents a Schwarzschild black hole immersed in an external purely gravitational field or, by analytic continuation of the Bertotti--Robinson parameter, a Schwarzschild black hole inside an expanding bubble of nothing.

Finally, motivated by the fact that the physical meaning of the $r_0$ parameter is to determine whether the aligned component of the electromagnetic field is independent of the non-aligned one, we have formally proved that, in these vacuum subcases, which therefore do not possess an electromagnetic field, the $r_0$ parameter is actually inessential and can be added or removed via a coordinate transformation. Hence, the two $r_0$ branches of the vacuum solution actually represent the same spacetime, while each could still provide a more convenient coordinate representation depending on the specific calculations.

As future extensions, it would be interesting to also introduce additional parameters into this class of solutions, such as the swirling or NUT parameter, and to investigate whether it is possible to obtain a parameterization of the more general Ovcharenko--Podolsk\'{y} class of charged and rotating black holes with an external Bertotti--Robinson electromagnetic field, such that all the subcases correspond to black holes with positive mass.
\acknowledgments
I would like to thank Marco Astorino, Sergio Luigi Cacciatori and Adriano Vigan\`{o} for useful discussions and valuable comments on the topics of this paper.
This work was partly supported by INFN.

\end{document}